\documentclass{emulateapj}
\usepackage{graphicx,times}
\slugcomment{Received 2007 August 6; accepted 2007 September 23}

\shorttitle{Millisecond pulsation in Aql X-1}
\shortauthors{P.Casella et al.}
\begin{document}
\def\new#1{{#1}}
\title{Discovery of coherent millisecond X-ray pulsations in Aql X-1}

\author{P. Casella\altaffilmark{1}, D. Altamirano\altaffilmark{1},
A. Patruno\altaffilmark{1}, R. Wijnands\altaffilmark{1}, and M. van der Klis\altaffilmark{1}}
\altaffiltext{1}{Astronomical Institute, ``Anton Pannekoek'', University of Amsterdam, Kruislaan 403, 1098 SJ Amsterdam, The Netherlands. E-mail: casella@science.uva.nl}
\thispagestyle{empty}

\begin{abstract}

We report the discovery of an episode of coherent millisecond X-ray
pulsation in the neutron star low-mass X-ray binary Aql X-1. The
episode lasts for slightly more than 150 seconds, during which the
pulse frequency is consistent with being constant. No X-ray burst or
other evidence of thermonuclear burning activity is seen in
correspondence with the pulsation, which can thus be identified as
occurring in the persistent emission. The pulsation frequency is
550.27 Hz, very close (0.5 Hz higher) to the maximum reported
frequency from burst oscillations in this source. Hence we identify
this frequency with the neutron star spin frequency. The pulsed
fraction is strongly energy dependent, ranging from $<$1\% at 3-5 keV
to $>$10\% at 16-30 keV. We discuss possible physical interpretations
and their consequences for our understanding of the lack of pulsation
in most neutron star low-mass X-ray binaries. If interpreted as
accretion-powered pulsation, Aql X-1 might play a key role in
understanding the differences between pulsating and non-pulsating
sources.

\end{abstract}
   \keywords{pulsars: individual
   (Aql X-1) --- stars: neutron --- X-rays: binaries }



\section{Introduction}

Accretion-powered millisecond X-ray pulsars (hereinafter AMSPs) had
been predicted in the early 1980s as the progenitors of millisecond
radio pulsars \citep{backeretal82,alparetal82}. The first
observational indication that neutron stars in low-mass X-ray binaries
(LMXBs) rotate rapidly came in 1996 with the discovery of slightly
drifting in frequency millisecond oscillations during thermonuclear
X-ray bursts \citep[for a review see][]{strobild06}. However it was
not until 1998 that the first AMSP was discovered
\citep{wvdk98}. Since then, a total of eight AMSPs have been found out
of the $>150$ LMXBs known up to date \citep{liuetal07}.

Since the theoretical prediction of the existence of AMSPs was made,
the main issue remained to explain the lack of pulsation in the
persistent X-ray emission of the majority of LMXBs. In recent decades
many theoretical efforts have been made to explain this, the main
question remaining whether the pulsation is hidden from the observer
or not produced at all. At present, the scenarios most often
considered are: (a) the magnetic field in non-pulsating LMXBs is too
weak to allow channeling of the matter onto the magnetic poles;
(b) the magnetic field has comparable strength inside most LMXB
neutron stars, but in the large majority of them it has been
``buried'' by accretion, resulting in a very low surface magnetic
field \citep[e.g.,][]{cummingetal01} which again is too weak to allow
channeling of matter. After the accretion stops, the magnetic field
eventually emerges from the neutron star surface and assumes its
intrinsic value; (c) pulsations {\it are} produced in all LMXBs, but
in most of them they are attenuated by a surrounding scattering medium
that washes out the coherent beamed pulsation
\citep{brainerdlamb87,kylafis87,titarchuketal02}; (d) the pulsations
are attenuated by gravitational lensing from the neutron star
\citep[e.g.,][]{meszarosetal88}.

From an observational point of view, since the discovery of the first
AMSP efforts have been focused on finding differences between the
sources showing pulsation and those that do not, in order to test
different theoretical hypotheses. Possible observed differences so far
are the orbital period \citep[which is on average shorter in AMSPs
than in other LMXBs, see e.g.,][]{kaaretetal06} and the time-averaged
accretion rate \citep[which is considered to be on average smaller in
AMSPs than in other LMXBs, see e.g.,][]{galloway06}.  However, 
although it is probable that orbital period and time-averaged
accretion rate play an important role in the determination of AMSP
properties, the reason for the lack of pulsation in most of LMXBs
still has to be found.

The properties of the seventh discovered AMSP \citep[HETE
J1900.1-2455,][]{kaaretetal06,gallowayetal07} gave new insights to
this issue. This source has a much higher inferred time-averaged
accretion rate than in the other AMSPs. The pulsation became
undetectable after two months from the beginning of the outburst, at
strong variance with the other AMSPs in which pulsations were always
detectable until the end of the outbursts. This has been interpreted
as evidence for burying of the magnetic field by the accreted material
during the outburst \citep{gallowayetal07}.

Transient highly coherent pulsations were also observed in 4U 1636-53
\citep{stromark02} during an $\sim$800 s interval. However, the
pulsations were detected at the flux maximum of a superburst, hence
they were likely nuclear-powered. \new{This is different compared to
the seven AMSP, where pulsations were interpreted in terms of an
hotspot resulting from magnetic channeling of matter onto the neutron
star surface.}

We are searching the full Rossi X-ray Timing Explorer (RXTE) archive
data for coherent pulsations \citep[see also][]{altamiranoetal07}. 
In this letter we report the
discovery of an episode of coherent millisecond X-ray pulsarions in
the neutron star binary Aql X-1. The LMXB recurrent X-ray transient
Aql X-1 is an atoll source \citep{reigetal00} wich shows kHz
quasi-periodic oscillations, X-ray bursts and burst oscillations
\citep{zhangetal98b}. An orbital period of 18.95 hr was obtained by
optical measurments \citep{chevalier91,welshetal00}. The presumed
onset of the propeller effect \citep{illarionov75,stellaetal86}
allowed \citet{campanaetal98} to estimate a value for the magnetic
field of $1-3\times10^8 ({D\over2.5~kpc})$ G, consistent with the
magnetic field expected (and later on measured) in AMSPs \cite[see
also][]{zhangetal98a}.

We analyzed RXTE archive data of Aql X-1 and found that the source
showed coherent pulsation at a frequency near the neutron star spin
frequency (as inferred from burst oscillations) in its persistent
emission for 150 seconds during the peak of its 1998 outburst. The
discovery of this pulsation episode may provide new insights on the
issue of why some LMXBs pulse and some do not, after many years of
debate.


\section{Data Analysis} \label{dataanalysis}

   \begin{figure}
     \plotone{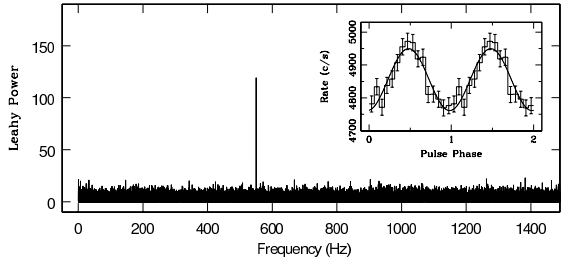}
     \caption{Power spectrum of the last 128s data interval of
     observation 30188-03-05-00 (see Fig. \ref{lc}). The power
     spectrum was obtained from the full bandpass event mode data. The
     frequency resolution is 1/128 Hz. The pulse is clearly visible at
     $\sim$550 Hz. In the inset we plot the 2-60 keV light curve
     folded at the pulsation period. Two cycles are plotted for
     clarity. The solid line shows the best sinusoidal fit.}
     \label{pds}
   \end{figure}


We analyzed the whole PCA public archive of Aql X-1, searching
for coherent pulsation in its persistent emission. A total of 363
observations were analysed. In addition to the two Standard Modes
always present, the PCA data in different observations were obtained
in two different high-time resolution modes (either Event Mode
(125$\mu$s) or GoodXenon($\sim1\mu$s)); to all data we applied
barycenter correction for Earth and satellite motion. For each
observation we computed fast Fourier transforms from 128s data
intervals (after filtering for the presence of X-ray bursts) and
obtained power spectra with a Nyquist frequency of 4096 Hz and a
frequency resolution of 1/128 Hz. We searched all power spectra for
significant power at frequencies close to the expected pulse frequency
($\sim$550 Hz). Only one detection was found with a Leahy power of 120
\citep{leahyetal83}, corresponding to a single trial chance
probability of $9\times10^{-27}$ at a frequency of 550.27 Hz (see
Fig. \ref{pds}).

The single trial significance of the pulsation is 11 $\sigma$. Once we
take into account the number of trials (i.e. the number of frequency
bins times the number of 128 s power spectra we searched) the
significance becomes 9 $\sigma$ ($3\times10^{-17}$ chance
probability). Combined with its frequency being near that of the
previously observed burst oscillations, we conclude that the pulsation
is real and intrinsic to Aql X-1 and represents the spin of the
neutron star in this source. A full description of the total data
analysis is beyond the scope of this letter and will be the subject of
a forthcoming publication. Here we focus on the detailed analysis
performed of the single observation where pulsation was discovered.

\begin{figure} 
     \plotone{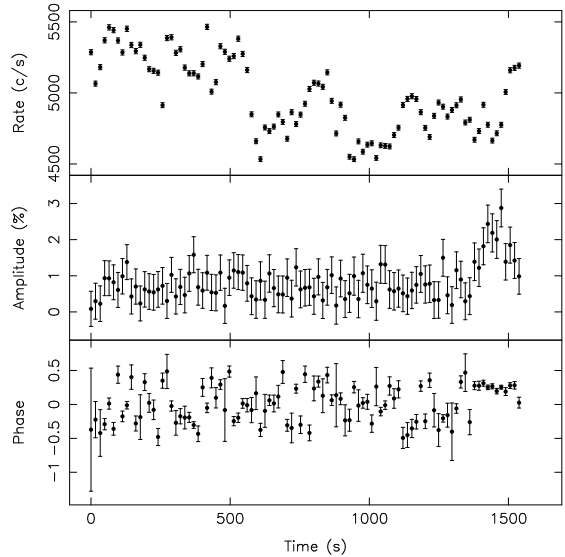}
     \caption{{\it Top panel}: 2-60 light curve with 16 seconds bin
      size.  {\it Bottom panels}: Amplitude ({\it middle}) and phase
      ({\it bottom}) by epoch folding technique every 16 seconds,
      vs. time. Values of amplitude and phase between 0 and $\sim$1400 s are
      consistent with those expected from Poissonian noise.}
     \label{lc}
   \end{figure}

The pulsation was detected at the end of a 1600-second long
observation starting on 1998 March 10 at 22:28 UT (ObsId:
30188-03-05-00).  All five proportional counter units were active
during the whole observation, with a total 2-60 keV count rate varying
between 4500 and 5500 counts/sec. During this observation high-time
resolution data were available in Event Mode with 125$\mu$s time
resolution and 64 energy channels over the 2-60 keV instrument
bandpass. The light curve of the observation is shown in the top panel
of Figure \ref{lc}.

The high power was detected in the last 128s interval of the
observation. By using an epoch folding technique we obtain a period
estimate of 1.8172746 ms. The folded profile was well modeled with a
sinusoidal component with 1.9$\pm$0.2 \% amplitude (see inset of
Fig. \ref{pds}). No second harmonic could be detected, with a 95\%
upper limit on the amplitude of 0.8 \%. To study the evolution and the
properties of the signal we extracted a light curve with maximum time
resolution, divided it into 16-second segments and determined the
pulsation amplitude and phase in each segment by an epoch folding
technique (see Fig. \ref{lc}). The pulsation is detected for $\sim150$
seconds at the end of the observation; its amplitude gradually
increases and then decreases again over these 150 seconds. To estimate
the accuracy of our period measurement we performed a linear fit to
the phases in the interval where pulsations are detected. Over this
interval the phase is consistent with being constant. From a 3
$\sigma$ upper limit of 10$^{-3}$ s$^{-1}$ on the slope, we derive a
best period measurement of 1.817275(3) ms. The short duration of the
pulsation did not allow us to obtain any useful upper limit on the
frequency drift. From the number of cycles we were able to phase
connect, we can estimate the coherence of the pulsation as Q $\ge
8\times10^4$. \new{No correction for the binary orbital motion was
possible, given the small expected drift ($<$ 2 mHz in 150 seconds)
and the large uncertainity on the orbital phase \citep[$\sim$80\%,
applying the orbital solution from][]{welshetal00}}.

\begin{figure}
     \plotone{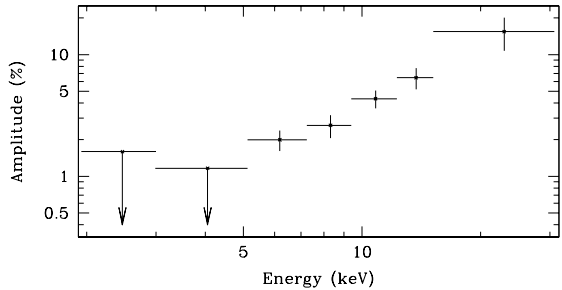}
     \caption{Amplitude of the pulsation as a function of energy.
              1 $\sigma$ errors are shown.}
      \label{ampene}
   \end{figure}

In order to study the energy dependence of the pulsation we extracted
light curves with maximum time resolution in different energy ranges
and determined the pulsation amplitude and phase in each of them. In
Figure \ref{ampene}, we show the amplitude vs. energy diagram. The
pulsations have a strong energy dependence, being not detected at low
energies (with an upper limit on its amplitude of $\sim$1\% below 5
keV) and strong at high energies.

\section{Discussion} \label{discussion}

We discovered millisecond X-ray pulsations in the persistent X-ray
emission of the LMXB Aql X-1. The pulsation is transient, present for
only $\sim$150 seconds and has an average fractional amplitude (over
the full 2-60 keV instrumental bandpass) of $\sim$2\%, increasing up
to $>$10\% at energies above $\sim$16 keV. It appears and disappears
gradually, on a time scale of a few tens of seconds, with a 3
$\sigma$ upper limit on its amplitude of 0.7\% during the $\sim$128
seconds before its appearance.

This pulsating episode appears to be unique in the whole RXTE archive
of Aql X-1 observations. In one third of the observations
(corresponding to an exposure of 500 ks) the count rate is high enough
to allow the detection (3-$\sigma$, single trial) of a similar pulsating
episode as the one we observed. Even considering only these data
(instead of the total analysed exposure of 1300 ks) we obtain a
recurrence rate smaller than $3\times10^{-4}$. Such extreme rarety
in itself is extremely informative. Any physical interpretation must
not only in fact explain the appearance of the pulsation, but also its
extremely low occurrence rate.

The frequency of the pulsation is 550.27 Hz, $\sim$0.53 Hz higher than
the reported asymptotic frequencies for burst oscillations in this
source \new{ \citep{zhangetal98b}}. The light curve shows no evidence of an
X-ray burst or other obvious feature within 1500 seconds before the
appearance of the pulsation. Together with its high coherence, this
leads us to conclude that the observed pulsation arises from a hot
spot spinning with the neutron star. This makes Aql X-1 the
millisecond X-ray pulsar with the longest orbital period and with the
highest time-average accretion rate known at present.

The detection of this fast appearing pulsation in a LMXBs leads once
again to the same question: is the pulsation in Aql X-1 usually hidden
from the observer, by some scattering or screening medium that very
rarely disappears (or strongly reduces its otpical thickness) for
$\sim$150 seconds? Or is the pulsation usually absent, the episode we
discovered being the result of an occasional and rare asymmetry
on the neutron star surface?

In the following we analyse different possible scenarios:

\subsection{Permanent pulsation} \label{discus:permanent}

If a hot spot is normally present on the neutron star surface, the
question is what made it observable only for $\sim$150 seconds.
Different authors have argued against or in favor of the hypothesis
that in non-pulsating sources the pulsation is washed out by some
optically thick scattering media. The main issue is the value of the
optical depth $\tau$ needed to attenuate the beamed
oscillations. \citet{psaltischakrabarty99} discuss the apparent
similarities between values of $\tau$ in pulsating and non-pulsating
sources, and discard this hypothesis. \citet{titarchuketal02} report
spectral fits giving other values of $\tau$ (for different
sources), and conclude that the data {\it do} support this
hypothesis. More recently, \citet{gogusetal07} report new,
independent spectral fits and conclude that in non-pulsating LMXBs
$\tau$ is not large enough to cause the pulsations to disappear
\citep[unless the electron temperature is very low, see
also][]{titarchuketal07}.

This issue appears to be difficult to resolve, mainly because of a
substantial degeneracy between optical depth and electron temperature
in the spectral fits. However, the pulsation episode in Aql X-1 now
allows us to provide new insights into this issue. Even though, because
of their degeneracy, the absolute values of spectral parameters can not
be properly constrained, one would expect them to change when the
pulsation appears. In particular, the appearance of the pulsation, if
caused by a temporary (total or local) decrease of the optical depth
of the scattering media, is expected to be accompanied by a decrease
of the comptonization parameter, hence by a softening of the energy
spectrum. However, our spectral fits show that a two-component model
(disk plus boundary layer) is sufficient to fit the data, without the need
for any additional comptonized component. Moreover, neither spectral
fitting nor analysis of hardness light curves show any evidence of
spectral variability associated with the appearance of the
pulsation. Furthermore, the similarities between the spectral
properties of Aql X-1 and many other LMXBs would lead one to expect
similar episodes of pulsation in most of LMXBs.  On the other hand,
the very hard spectrum of the observed pulsation (see
Fig. \ref{ampene}) suggests that the beamed radiation is strongly
comptonized \citep[for a discussion on the expected energy dependence
in the case of the screening scenario see][]{falanga07}.

An alternative could be to assume a non-isotropic screening medium: an
occasional and rare hole in the medium or a reflecting cloud could
have allowed the radiation from the hot spot to avoid the screening
medium, thus becoming visible. Finally, the hot spot itself could have
moved becoming temporarily visible, either because of a different
absorption through the line of sight or because of a change in the
rotational of gravitational lensing geometry. We note that all these
hypotheses could in principle be consistent with the observed gradual
increase and decrease of the pulsation amplitude, but require very
special and fine-tuned geometries.

\subsubsection{Estimate of the magnetic field} \label{discus:B}

Under the standard dipolar magnetic channeling hypothesis for creating the hot
spot, we can give an estimate of the magnetic field. The magnetic field
must be strong enough to locally disrupt the disk flow. The pulsation
episode happens when the source is in the soft state, close to the
peak of the outburst. The 2-10 keV flux is $\sim 8.8\times10^{-9}$ erg
s$^{-1}$ cm$^{-2}$ which corresponds, assuming a distance of 5 kpc
\citep{jonkernelemans04} and a bolometric correction of 1.4
\citep{migliarifender06}, to a bolometric luminosity of $\sim
3.7\times10^{37}$ erg s$^{-1}\sim0.15L_{\rm E}$.  Given the relatively
high luminosity, by imposing that the radius at which this happens is
larger than the neutron star radius we can derive a stringent lower
limit on the magnetic dipole moment \citep{psaltischakrabarty99} as $
\mu \ge 0.47 \times 10^{26}~\mbox{G~cm}^3\;$ which corresponds to a
magnetic field of $\sim 10^8$ G at the pole.

This lower limit is consistent with the upper limit of
$\sim1.8\times10^9$ G obtained by \citet{disalvoburderi03} from the
quiescent luminosity and with the estimate of $2\sim6\times10^8 $ G
obtained by \citet[assuming a corrected distance of 5 kpc instead of
2.5 kpc]{campanaetal98} from the reported onset of the propeller
effect. Furthermore, our lower limit rules out the hypothesis of
residual accretion onto the neutron star to explain the quiescent X-ray
luminosity in Aql X-1, for which an upper limit for the magnetic field of
$\sim4\times10^7$ G had been derived (hypothesis a1 of
\citet{disalvoburderi03}).

\subsection{Transient pulsation} \label{discus:transient}

The second possibility is that of a temporary asymmetry on the
neutron star surface. Let us examine possible causes:

\noindent {\it a) Magnetic channeling}

Accretion rate and magnetic field are the two key physical quantities
playing a role in channeling of matter.  A transient episode of
channeling of matter could arise due to a temporary change in the
accretion rate (although the X-ray flux, which is thought to be a good
indicator of $\dot{M}$, does not show any variability correlated with
the appearance of the pulsation) or a change in the accretion
geometry.

An alternative is the occurrence of some variability of the magnetic
field. In the ``burying'' scenario, timescales for the magnetic field
to emerge are set by the Ohmic diffusion time and are at least of the
order of 100-1000 yr \citep{cummingetal01}, obviously far larger than
the few tens of seconds observed in Aql X-1. However a local temporary
decrease of the Ohmic diffusion timescale could in principle let the
magnetic field emerge.  Possible causes could be a temporary local
disruption of the screening currents, a starquake, or magnetic
reconnections. Perhaps, a temporary local magnetic loop might form,
somewhat similar to those observed in the solar corona, where hot
coronal plasma ``populates'' different field lines. The magnetic field
in this loop might be {\it locally} strong enough to disturb the flow
of matter onto the neutron star surface.  The resulting hot spot (non
necessarily on the magnetic pole) creates the observed pulsation.
However multipolar magnetic fields in neutron stars have not been
studied in detail, and these phenomena have to our knowledge not been
theoretically predicted.

\noindent {\it b) Nuclear burning}

The hot spot might be associated to a confined nuclear burning event.
After its onset, nuclear burning is expected to spread over the neutron
star surface in a very short timescale. Hence one would not expect the
resulting pulsation to last more than a few seconds. However the burst
oscillations observed in many sources during the decay of X-ray bursts
strongly challenge this simplified picture. Moreover, the discovery of
a very coherent pulsation during a superburst in 4U 1636-53
\citep{stromark02} showed that long-lived, very coherent pulsations
might arise due to nuclear burning. However the lack of any other
X-ray signature of nulcear burning (such as an X-ray burst) associated
with the pulsation in Aql X-1 strongly argues against this hypothesis.

\noindent {\it \new{c) Surface wave}}

\new{An asymmetry on the neutron star surface might result from an
unstable surface mode, as the ones suggested to explain burst
oscillations \citep[see e.g.,][]{pirobildsten04,narayancooper07}. Since
the amplitudes of surface modes have been predicted to increase with
energy \citep{pirobildsten06}, the hard energy dependence observed in
Aql X-1, more similar to that of burst oscillations than that of
accretion-powered pulsations, supports this possibility.}

\noindent {\it d) Other possibilities}

The extreme rarity of this pulsation episode leads us to examine also
more exotic physical explanations. Possible origins for a transient
hot spot could be a self-luminous magnetic event on the neutron star
surface, as magnetic reconnections or annihilations, or a sudden
release of potential energy in the deep crust during a starquake.

Finally, we mention the possibility that some of the discussed
instabilities could arise due to a collision with an external body
\citep[see e.g.,][]{colgate81}. The fluence of the observed pulsation
is $\sim10^{35}$ergs, which corresponds to the free fall kinetic
energy of a mass of $\sim$15 km$^3$ of iron. However, the observed
gradual increase in the pulsation amplitude would be difficult to
explain within this scenario. The probability to have such an impact
is presumably extremely low. On the other hand, the observed event
{\it is} very rare.

\new{Whatever is the physical origin of the observed pulsations, their
discovery implies the possible occurrance of transient pulsations in
any of the {\it apparently} non-pulsating LMXBs.}

\acknowledgments

The authors would like to thank A. G{\"u}rkan, G. Israel, and
L. Stella for very useful comments and stimulating discussions. We
also thank L. Bildsten and R. Cooper for pointing out the surface mode
possibility.



\newpage

\end{document}